%% file: parallelspin_v5.tex
\documentclass[11pt,a4paper]{article}
\usepackage{jheppub}

\usepackage{amsmath,amssymb}
\usepackage{enumerate}
\usepackage{amsfonts}

\usepackage{subfigure}

\usepackage{allpurpose}

\usepackage{tikz}

\usepackage[marginal,perpage,symbol]{footmisc}

\usepackage{slashed} 

\graphicspath{{./figures/}}

\everymath{\displaystyle}
\makeatletter
\gdef\@fpheader{}
\makeatother

\usepackage[nottoc,notlof,notlot]{tocbibind}

\usepackage{indentfirst}

\begin{document}

\title{The precession of particle spin in spherical symmetric spacetimes}
\date{\today}

\author[1]{Xiankai Pang\note[1]{The corresponding author.},}
\emailAdd{xkpang@cwnu.edu.cn}
\author{Qingquan Jiang,}
\emailAdd{qqjiangphys@yeah.net}

\author{Yunchuan Xiang,}
\emailAdd{xiang{\_}yunchuan@yeah.net}

\author{and Gao-Ming Deng}
\emailAdd{gmd2014cwnu@126.com}

\affiliation{\normalfont \normalfont School of Physics and Astronomy, China West Normal University, Nanchong, 637009, China}

\abstract{
	In this work, we will explore the precession of particle spins in spherical spacetimes. We first argue that the geometrical optics (WKB) approximation is insufficient, due to the absence of a glory spot in the backward scattering of massless particles, making an analysis of spin precession necessary. We then derive the precession equation assuming the spin is parallel transported, which is supported by the sub-leading order of the WKB approximation. The precession equation applies to both massless and massive particles. For particles moving at the speed of light, we show that spin is always reversed after backward scattering in any spherically symmetric spacetime, confirming the absence of a glory spot for massless particles. Finally, we solve the precession equation for Schwarzschild and Reissner-Nordstr\"om spacetimes and discuss the spin precession of massive particles, particularly in the non-relativistic limit. We find that, in Schwarzschild spacetime, the spin precession for particles moving with very small velocities compared to the speed of light depends only on the deflection angle, while in Reissner-Nordstr\"om spacetime, it also depends on the black hole charge, as revealed by the expansion derived from the strong lensing approximation.
}


\arxivnumber{2410.04323}

\maketitle

\section{Introduction}

The bending of light rays in gravitation fields is expected even before the full establishment of \acrfull{gr}, as a direct consequence of the equivalence principle~\cite{Einstein:1911vc}, and the observation of the deflection~\cite{Dyson:1920cwa} constitutes an important pillar in showing \acrshort{gr} as the correct theory of gravity. Nowadays, the gravitational lensing of light has become an important tool in understanding our universe, such as mapping the distribution of dark matter~\cite{Kaiser:1992ps}, investigating the large scale structure~\cite{Bacon:2000sy}, and even taking photos of black holes~\cite{Akiyama:2019bqs}.

By taking light as rays, we resort to the \emph{geometrical optics} approximation (or WKB approximation in quantum mechanics) by letting the wavelength $\lambda$ go to zero~\cite{Landau:1982dva}, in this way, the wave nature is ignored and light can be seen as massless particles. This is justified since the length scale in the gravitational lensing (such as the black hole radius) is usually much larger than the wavelength. The approximation is not limited to light; it can also apply to other fields, such as scalar fields and Dirac fermions, whether massless or massive. One problem is, however, at the leading order of the WKB approximation, we not only ignore the wave nature of the fields, but also the internal degrees of freedom (such as spin\footnote{Rigorously speaking, for massless particles like photons the spin is not a well-defined quantity, and we should consider \emph{helicity} instead. But the distinction is not important for the current work, and we will stick to the term `spin' for convenience.}) of them. 

More specifically, no matter whether the field has spin or not, at the leading order, WKB approximation always provides us with the same geodesic equation~\cite{Landau:1982dva,Stone:2014fja,Oancea:2022utx}. For example, the scattering of light rays (with spin $s=1$) would be identical to the scattering of a massless scalar (with $s=0$) under the leading order WKB approximation. While this works for most scattering scenarios, it fails in the case of scattering in the backward direction~\cite{Crispino:2009xt,Futterman:1988sbh}.
Partial wave analysis shows that for massless scalar fields, the cross-section in the backward direction is diverging, leading to the so-called \emph{glory spot}\footnote{It's a bright spot in the backward direction, i.e., where deflection angle $\Delta\varphi=\pi$. See equation \eqref{eq:deltaphidef} for the definition of the deflection angle $\Delta\varphi$.}~\cite{Matzner:1985rjn,Crispino:2009kia,Macedo:2015qma,Huang:2020bdf}, whose intension can be estimated by \emph{glory approximation} \cite{Ford:1959sd,DeWittMorette:1984pk,Zhang:1984vt,Crispino:2009xt}.  For the particles with non-vanishing spin, however, the cross-section in the backward direction vanishes, and there is no glory spot \cite{Zhang:1984vt,Futterman:1988sbh,Dolan:2006vj,Crispino:2009xt}.
This discrepancy indicates that the leading-order WKB approximation is insufficient for describing the scattering of spinning particles.

\begin{figure}[htpb]
	\centering
	\subfigure[Circular orbits on photon sphere]{\label{fig:parallelsphere}\includegraphics[width=0.38\textwidth]{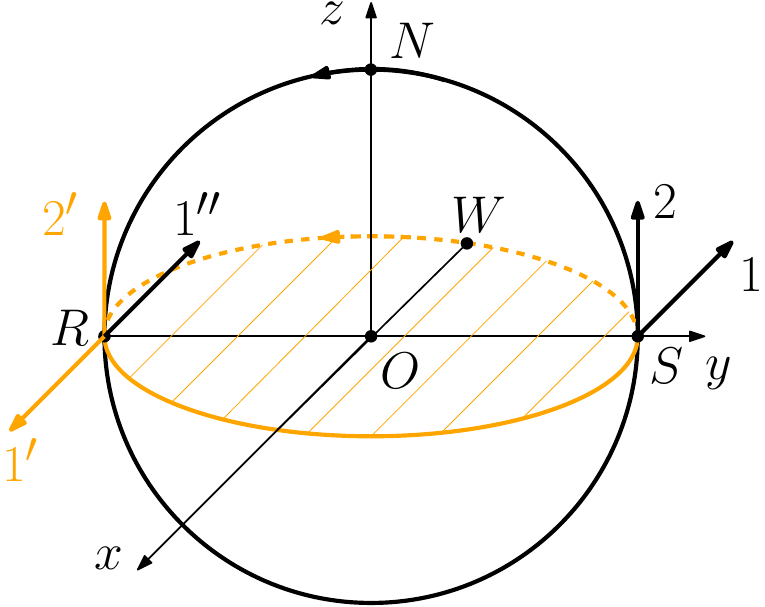}}\quad
	\subfigure[Unbounded orbits with backward scattering]{\label{fig:parallelunbounded}\raisebox{26pt}{\includegraphics[width=0.54\textwidth]{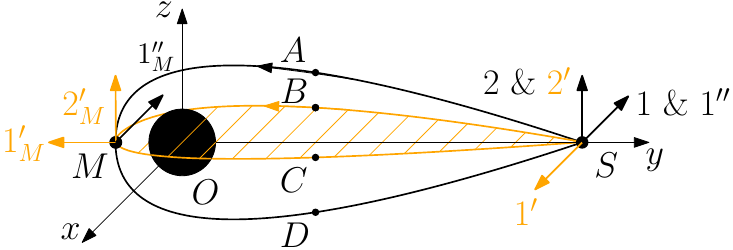}}}
	\caption{The spin precession of massless particles for different kinds of orbit. In both figures \ref{fig:parallelsphere} and \ref{fig:parallelunbounded}, the yellow orbits (SWR in \ref{fig:parallelsphere} and SBMCS in \ref{fig:parallelunbounded}) lie in the equatorial plane, while the black orbits (SNR in \ref{fig:parallelsphere} and SAMDS in \ref{fig:parallelunbounded}) lie in the perpendicular one. We will see that for a massless particle emitted from $S$, its spin vector $1$ becomes $1'$ or $1''$ after being parallel transported along the yellow or black orbits, respectively. $1'$ and $1''$ will cancel since they have opposite directions, resulting in the vanishing glory spot for massless particles. For spin vector $2$, we didn't show $2''$, the result after moving along black orbits, to avoid redundancy.}
	  \label{fig:parallelsphereunbounded}
\end{figure}

Classically, the absence of glory spot for massless spinning particles can be accounted by the precession of spins, as illustrated in figure~\ref{fig:parallelunbounded}, where the particle spin (represented as a vector) is parallel transported along the geodesic~\cite{Zhang:1984vt,Futterman:1988sbh,Crispino:2009xt}. We will show later that for a massless particle emitted and received at the same point $S$ (hence the particle is scattered to the backward direction), the spin vector $1$ will become $1'$ (whose direction is reversed) after the particle travels along the orbit SBMCS in the equatorial plane. At the same time, it becomes $1''$ (whose direction is the same as vector $1$) after travelling along the orbit SAMDS, sitting in the plane perpendicular to the equatorial one. Since vectors $1'$ and $1''$ have opposite directions they will cancel out. And in a spherical symmetric spacetime, for any orbit, there is always one in the perpendicular plane, such that after following these two orbits a particle will have opposite spins, and there will be a perfect cancellation~\cite{Crispino:2009xt}. 

In this paper, we are going to investigate the parallel transportation of particle spins along the geodesics, for both massless and massive particles. For our purpose, the spin of various fields can be characterized by a $4$-vector $S^\mu$ \cite{Dolan:2006vj,Pomeransky:2000pb,Khriplovich:2008ni}, which will be parallel transported along the geodesic~\cite{Weinberg:1972kfs}.
And it turns out the parallel transportation of spin will emerge at the sub-leading order of the WKB approximation~\cite{Poplawski:2009fb,Obukhov:2013zca,Oancea:2022utx}. 
In other words, by expanding the wave equation for small wavelength, at the leading order we will get the geodesic equation, and at the sub-leading order, we will get the precession of the particle spin~\cite{Stone:2014fja,Oancea:2022utx}. 

It's worth mentioning that the WKB approximation has its limitations, although we are already able to study the behaviour of spins at the sub-leading order.
The point is, since the geodesic motion always appears at the leading order, one won't be able to obtain the spin-orbit coupling under the approximation~\cite{Rubinow:1963asd,Stone:2014fja,Oancea:2022utx}. 
To get the spin-dependent trajectories,  one may start with an extended object that subjects to the tidal force, and arrive at 
the \acrfull{mpeq} equations, where the particle trajectory will depend on the spin and deviates from the geodesics~\cite{Mathisson:2010opl,Papapetrou:1951pa,Barducci:1976wc,Rudiger:1981uu,Cianfrani:2008ug,Plyatsko:2015bia}, and the spin itself will also precess~\cite{Pomeransky:2000pb,Iorio:2011ubn,Ruangsri:2015cvg,Chakraborty:2016mhx}. The deviation affects both unbounded and bounded orbits. The former is responsible for the gravitational lensing and the scattering cross-section~\cite{Zhang:2022rnn,Jiang:2023yvo,Ying:2024wop}, while the latter one considers the circular orbits~\cite{Mohseni:2010rm,Jefremov:2015gza,Zhang:2017nhl,Zhang:2018eau,Zhang:2020qew} and periapsis shift~\cite{He:2023joa}. 
An analytical solution for the \acrshort{mpeq} equations has been found recently for massive spinning particles in spherical symmetric spacetimes~\cite{Witzany:2023bmq}. For massless particles, the \acrshort{mpeq} equations behaves differently \cite{Saturnini:1976mps,Duval:2018hzh,Duval:2016hxo}, and one has to use appropriate spin supplementary conditions~\cite{Poisson:2011nh,Deriglazov:2021gwa,Deriglazov:2022cvu,Frolov:2020uhn,Harte:2022dpo}. Various attempts in going beyond WKB approximation for massless particles have been taken~\cite{Harte:2018wni,Frolov:2024ebe,Frolov:2024qow,Frolov:2024olb}, and the gravitational Faraday and spin-Hall effects are also addressed~\cite{Shoom:2020zhr,Oancea:2020khc,Oancea:2023ylb}. The massless limit can also generalize to other spins, such as linear gravitational waves~\cite{Andersson:2020gsj} with spin $1$ and massless Dirac fields~\cite{Stone:2014fja,Oancea:2022utx} with spin $1/2$.

As we can see, however, the parallel transportation of spin along geodesics is already sufficient to get a vanishing glory spot in the backward direction for massless particles, and it's not necessary to dive into the more complex situations right now, when a detailed discussion of the backward scattering for massive particles, especially for the non-relativistic case where the particle speed is very small compared to the speed of light, is still missing. Therefore, in this manuscript, we will focus on the leading and sub-leading order of the WKB approximation, by considering only the parallel transportation of spins along geodesics for both massless and massive particles and leaving the possible extension to \acrshort{mpeq} equations for future works.

The manuscript is organized as follows. In section~\ref{sec:motionspin} we will derive the equations governing the precession of the particle spin, by letting the spin be parallel transported along the geodesic. Then in section~\ref{sec:circularorbits} we consider solving the precession equations for the simplest orbits, i.e., the circular ones, especially on the photon sphere. Section \ref{sec:unboundedorbits} shows how to solve the precession equation for more general orbits by changing variables. The solution allows us to define an angle to characterize the precession of the spin vector with respect to the initial direction in section~\ref{sec:anglexidef}, where we also show that the spin precession of ultra-relativistic particles is the same for any spherical symmetric spacetime. This results in the vanishing of the glory spot in backward scattering. In section~\ref{sec:examplesSchRN}, we apply these arguments to Schwarzschild and \acrfull{rn} spacetimes, where the non-relativistic limit is also under detailed investigation. Finally, in section~\ref{sec:summary} we discuss our results and point out possible extensions of the current work. 

For simplicity, without further specification we will use the natural unit where $G=c=1$, for the Newton constant $G$ and the speed of light $c$. 

\section{The precession of spin}
\label{sec:motionspin}
Generally, the metric in a spherically symmetric spacetime under the coordinates $x^\mu=(t,~r,~\theta,~\varphi)$ can be written as
\begin{IEEEeqnarray}{rCl}
	\dd s^2 &=& f(r)\dd t^2-\frac{1}{f(r)}\dd r^2 -r^2\dd\theta^2-r^2\sin^2\theta\dd\varphi^2 \quad,  \label{eq:frmetric}
\end{IEEEeqnarray} 
for an arbitrary function $f(r)$.
The spinning particle can be described by the angular momentum $4$-vector~\cite{Weinberg:1972kfs,Dolan:2006vj,Khriplovich:2008ni}
\begin{IEEEeqnarray}{rCl}
	\vec{S} &=& S^t\vec{e}_t + S^r\vec{e}_r +S^\theta\vec{e}_\theta +S^\varphi\vec{e}_\varphi\quad, 
\end{IEEEeqnarray} 
where $S^\mu=(S^t,S^r,S^\theta,S^\varphi)$ are components of the spin relative to the orthonormal basis spanned by $\vec{e}_t,~\vec{e}_r,~\vec{e}_\theta,~\vec{e}_\varphi$. These vectors are natural basis for each coordinate, with $\vec{e}_r=\frac{\partial}{\partial r}$, lies in the direction of increasing radial coordinate $r$, and similarly for the other three basis $\vec{e}_t,~\vec{e}_\theta$ and $\vec{e}_\varphi$.

The spin vector is orthogonal to the  $4$-velocity $\displaystyle U^\mu=\frac{\dd x^\mu}{\dd\tau}$ ($\tau$ is the proper time)\footnote{The reason why $S^\mu$ is orthogonal to $U^\mu$ is that in the rest frame of the particle we have $U^t=1$ and all other three components of the four-velocity vanish, while $S^\mu$ coincides with the $3$ angular momentum vector and $S^t=0$, so $S^\mu U_\mu=0$ in the rest frame~\cite{Weinberg:1972kfs}. Furthermore, the quantity $S^\mu U_\mu$ is a scalar, so the equality is valid in all frames.} 
\begin{IEEEeqnarray}{rCl}
	S^\mu U_\mu &=& 0  \quad, \label{eq:smuUmuorthogonal}
\end{IEEEeqnarray} 
For our purpose, it's sufficient to stick to the sub-leading order WKB approximation of field equations and ignore the spin-orbit coupling~\cite{Stone:2014fja,Oancea:2022utx}. Therefore, the particle trajectory will be a geodesic, and $S^\mu$ will be parallel transported along it
\begin{IEEEeqnarray}{rCl}
	\frac{\mathrm{D}S^\mu}{\dd\tau} &=&\frac{\dd S^\mu}{\dd\tau}+\Gamma^\mu_{\nu\lambda}S^\nu \frac{\dd x^\lambda}{\dd \tau} =0\quad,  \label{eq:smuparallel}
\end{IEEEeqnarray} 
where $\mathrm{D}$ represents the covariant differential and  $\Gamma^\mu_{\nu\lambda}$ are Christoffel symbols.

At the classical level, the spin of both photons and fermions can be described by the $4$-vector $S^\mu$ \cite{Dolan:2006vj,Deriglazov:2021gwa}. For the photon, the spin can be constructed using the polarization of electromagnetic tensor~\cite{Dolan:2017zgu,Deriglazov:2021gwa,Deriglazov:2022cvu}, 
or one can set $S^\mu$ as the normalized amplitude of the electromagnetic potential~\cite{Straumann:1984grr}. Either way $S^\mu$ will be parallel transported along the trajectory. 
For Dirac fermion $\psi$, on the other hand, the spin $4$-vector can be defined using~\cite{Dolan:2006vj,Cianfrani:2008ug}
\begin{IEEEeqnarray}{rCl}
	S^\mu &=& \bar{\psi}\gamma^\mu \gamma_5\psi \quad, \label{eq:smudiracdef}
\end{IEEEeqnarray} 
where $\gamma^\mu=e_a^{~\mu}\gamma^a$ is the position-dependent $\gamma$-matrices, and $e_a^{~\mu}$ represents the tetrad frame that is necessary to incorporate fermions in \acrshort{gr} \cite{Weinberg:1972kfs,Poplawski:2009fb}. For details see appendix~\ref{sec:paralleldiracspinor}. 

In appendix~\ref{sec:paralleldiracspinor}, we will show that for the latter case, $S^\mu$ will be parallel transported if the Dirac fermion does. More explicitly, we will show that under the definition~\eqref{eq:smudiracdef}, the parallel transportation equation~\eqref{eq:smuparallel} is valid if the Dirac fermion is parallel transported according to equation~\eqref{eq:DpsiGammamu}. Therefore, classically we can use the same transportation equation~\eqref{eq:smuparallel} to describe the precession of spins of photons and Dirac fermions as well as many other particles.

\subsection{The geodesic equation}

For convenience, we recall the geodesic equation in curved spacetime has the form
\begin{IEEEeqnarray}{rCl}
\frac{\dd^2 x^\mu}{\dd\tau^2}+\Gamma^\mu_{\nu\lambda}\frac{\dd x^\nu}{\dd\tau}\frac{\dd x^\lambda}{\dd\tau} &=&0  \quad. \label{eq:xmugeodesic}
\end{IEEEeqnarray} 
When the spacetime is spherically symmetric, a given geodesic will lie in the equatorial plane for all proper time $\tau$, on which we can choose $\theta=\frac{\pi}{2}$. Substituting the metric~\eqref{eq:frmetric} on the equatorial plane, the geodesic equation~\eqref{eq:xmugeodesic} becomes (where we have omitted the obvious one $\ddot{\theta}=0$)
\begin{IEEEeqnarray*}{rCl}
	 &&\frac{\dd f}{\dd r}\dot{r}\dot{t}+f\ddot{t}=0  \quad, \quad    f(r)\left\{f(r)\left[\frac{\dd f}{\dd r}\dot{t}^2-2r\dot{\varphi}^2\right]+2\ddot{r}\right\}- \frac{\dd f}{\dd r} \dot{r}^2=0 \quad ,  \quad 2\dot{r}\dot{\varphi}+r\ddot{\varphi}=0\quad,
\end{IEEEeqnarray*} 
where the dot $\dot{}$ denotes the derivative respect to the proper time $\tau$.
The geodesic equation can be integrated once such that  
\begin{IEEEeqnarray}{rCl}
	\dot{t} &=& \frac{E}{f(r)} \quad, \quad \dot{r}	= \sqrt{E^2-V_{\text{eff}}(r)} \quad,\quad \dot{\varphi} = \frac{L}{r^2} \quad. \label{eq:dottrthetaphi}
\end{IEEEeqnarray} 
In equation \eqref{eq:dottrthetaphi} we have introduced two integral constants, $E$ and $L$, to represent the energy and orbit angular momentum of the particle, respectively. The effective potential has the form 
\begin{IEEEeqnarray}{rCl}
	V_{\text{eff}}(r) &=& f(r)\left(\frac{L^2}{r^2}+\kappa\right) \quad, \label{eq:Veffdef}
\end{IEEEeqnarray} 
where $\kappa$ is the normalization of the $4$-velocity $U^\mu U_\mu=\kappa$. Under the metric \eqref{eq:frmetric}, we have $\kappa=1$ for massive particles while $\kappa=0$ for massless particles.

For unbounded orbits, it's more convenient to introduce velocity $v$ and impact parameter $b$ at infinity to replace the energy $E$ and angular momentum $L$~\cite{Pang:2018jpm}
\begin{IEEEeqnarray}{rCl}
	E &=& \frac{1}{\sqrt{1-v^2}} \quad, \quad L=\frac{bv}{\sqrt{1-v^2}}=b E v  \quad. \label{eq:ELbv}
\end{IEEEeqnarray} 
For massless particles, $v=1$ but $L/E=bv=b$ still holds.

Combining $\dot{r}$ and $\dot{\varphi}$ in equation~\eqref{eq:dottrthetaphi}, we obtain the orbit equation 
\begin{IEEEeqnarray}{rCl}
	\frac{\dd\varphi}{\dd r} &=& \frac{L}{r^2}\frac{1}{\sqrt{E^2-V_{\text{eff}}(r)}} \quad, \label{eq:orbitequationphir}
\end{IEEEeqnarray} 
which can be used to get the deflection angle for unbounded orbits, where the particle is lensed by the central black hole~\cite{Pang:2018jpm}. In fact, for the particles falling from $r=\infty$, the relation $\varphi=\varphi(r)$ can be given by 
\begin{IEEEeqnarray}{rCl}
	\varphi(r) &=& \int_{r}^{\infty} \frac{L}{x^2}\frac{\dd x}{\sqrt{E^2-V_{\text{eff}}(x)}} \quad. \label{eq:varphirsolint}
\end{IEEEeqnarray} 
For a particle coming from infinity and then being scattered to infinity again, the deflection angle $\Delta\varphi$ can be defined as~\cite{Bozza:2002zj,Pang:2018jpm} 
\begin{IEEEeqnarray}{rCl}
	\Delta\varphi &=& 2\varphi_{0}-\pi \quad, \label{eq:deltaphidef}
\end{IEEEeqnarray} 
where $\varphi_{0}=\varphi(r_0)$, and $r_0$ is the smallest distance from the orbit to the black hole center, as illustrated in figure~\ref{fig:anglephichixi}.

\subsection{The precession equation}

Substituting equation~\eqref{eq:dottrthetaphi} and metric~\eqref{eq:frmetric} into the parallel transport equation~\eqref{eq:smuparallel} for spin $S^\mu$, after constraining on the equatorial plane $\theta=\frac{\pi}{2}$, we obtain four equations of $S^t,~S^r,~S^\theta$ and $S^\varphi$, respectively
\begin{IEEEeqnarray}{rCl}
	&&\sqrt{E^2-V_{\text{eff}}(r)} \frac{\dd f}{\dd r}S^t+\frac{E }{f(r)}\frac{\dd f}{\dd r}S^r+2f(r)\dot{S}^t =0   \quad,\IEEEyesnumber \IEEEyessubnumber \label{eq:stdot} \\ 
	&&f(r)\left(2\dot{S}^r+E \frac{\dd f}{\dd r}S^t -\frac{2Lf(r)}{r}S^\varphi\right) - \sqrt{E^2-V_{\text{eff}}(r)} \frac{\dd f}{\dd r}S^r= 0  \quad, \IEEEyessubnumber \label{eq:srdot} \\
	&&\sqrt{E^2-V_{\text{eff}}(r)}S^\theta +r \dot{S}^\theta=0\quad, \IEEEyessubnumber \label{eq:sthetadot}\\ 
	&& \frac{L }{r^2}S^r+\sqrt{E^2-V_{\text{eff}}(r)}S^\varphi +r\dot{S}^\varphi=0 \quad. \IEEEyessubnumber \label{eq:sphidot}
\end{IEEEeqnarray} 
Recall equation~\eqref{eq:dottrthetaphi} we see that the third equation~\eqref{eq:sthetadot} is simply 
\begin{IEEEeqnarray*}{rCl}
	\dot{r}S^\theta+r\dot{S}^\theta &=& \frac{\dd}{\dd\tau}(rS^\theta)=0 \quad, 
\end{IEEEeqnarray*} 
which can be solved by 
\begin{IEEEeqnarray}{rCl}
	S^\theta(r) &=& \frac{S^\theta_0}{r} \quad, 
\end{IEEEeqnarray} 
where $S^\theta_0$ is determined by initial conditions. Therefore, for the circular orbits (where $r$ is constant) in the equatorial plane $\theta=\frac{\pi}{2}$, the $S^\theta$ component doesn't change, which justifies figure~\ref{fig:parallelsphere} and~\ref{fig:parallelunbounded}, for the vector $2$ moving along the path SWR (in figure~\ref{fig:parallelsphere}) or SBMCS (in figure~\ref{fig:parallelunbounded}) to become $2'$. Due to the spherical symmetry and the fact that paths SNR in figure~\ref{fig:parallelsphere} and SAMDS in figure~\ref{fig:parallelunbounded} are perpendicular to the equatorial plane, the direction of spin vector $1$, which lie in the equatorial plane, remain unchanged after the parallel transportation along these two paths.

For the remaining three equations~\eqref{eq:stdot},~\eqref{eq:srdot} and~\eqref{eq:sphidot}, we can eliminate $S^t$ with the help of the orthogonal relation~\eqref{eq:smuUmuorthogonal}
\begin{IEEEeqnarray*}{rCl}
	&&2\dot{S}^r+L \frac{\dd f}{\dd r}S^\varphi-\frac{2Lf}{r}S^\varphi =0  \quad,\\ 
	&& r \dot{S}^\varphi+\sqrt{E^2-V_{\text{eff}}}S^\varphi+\frac{L}{r^2}S^r =0 \quad. 
\end{IEEEeqnarray*}
Therefore, in the following, we can only consider components $S^r$ and $\bar{S}^\varphi$.
To further simplify these two equations, we can change the variables to $\varphi$ and introduce $\bar{S}^\varphi=r S^\varphi$ to obtain~\cite{Dolan:2006vj}
\begin{IEEEeqnarray}{rCl}
	 && \frac{\dd}{\dd\varphi} S^r(\varphi)-\left[f(r)-\frac{r}{2}\frac{\dd f}{\dd r}\right]\bar{S}^\varphi(\varphi) =0  \quad,  \IEEEyesnumber \IEEEyessubnumber \label{eq:srpphi}\\ 
	 &&\frac{\dd }{\dd\varphi}\bar{S}^\varphi(\varphi)+S^r(\varphi)=0 \quad . \IEEEyessubnumber \label{eq:sphipphi}
\end{IEEEeqnarray} 
And we can eliminate $S^r$ by taking a further derivative to equation~\eqref{eq:sphipphi} respect to $\varphi$, which is the precession equation we are looking for 
\begin{IEEEeqnarray}{rCl}
	\frac{\dd^2}{\dd\varphi^2}\bar{S}^\varphi + \left[f(r)-\frac{r}{2}\frac{\dd f}{\dd r}\right]\bar{S}^\varphi(\varphi)&=&0  \quad. \label{eq:sphippphi}
\end{IEEEeqnarray} 

Note that in the precession equations of spin, for example equation~\eqref{eq:sphippphi}, $r$ is also a function of $\varphi$. In other words, the solution of equation~\eqref{eq:sphippphi} would depend on the orbit $r=r(\varphi)$. In the following two sections, we will consider circular orbits $r=r_0$ and unbounded orbits, respectively.

Before we close this section, however, we would like to emphasize that the quantities $S^r$ and $\bar{S}^\varphi$ we have introduced are components with respect to the natural basis $\vec{e}_r$ and $\vec{e}_\varphi$, as we have mentioned 
\begin{IEEEeqnarray}{rCl}
	\vec{S} &=& S^r\vec{e}_r+\bar{S}^\varphi \vec{e}_\varphi \quad. 
\end{IEEEeqnarray} 
In particular, the constant components $S^r$ and $\bar{S}^\varphi$ represent a spin vector that rotates along the orthonormal frame spanned by $\vec{e}_r$ and $\vec{e}_\varphi$, as we will see in the next section for circular orbits on the photon sphere.

\section{Circular orbits and the photon sphere}
\label{sec:circularorbits}

Before discussing the more involved situations, we consider the circular orbit $r=r_0$, especially the one on the photon sphere~\cite{Virbhadra:1999nm,Claudel:2000yi,Virbhadra:2002ju,Iyer:2006cn}. 
Let's recall that for circular orbits, we would require~\cite{Jefremov:2015gza,Jia:2017nen}
\begin{IEEEeqnarray*}{rCl}
	E^2-V_{\text{eff}}&=& 0 \quad,   \quad \text{and} \quad \frac{\dd V_{\text{eff}}}{\dd r}  =0
\end{IEEEeqnarray*} 
For massless particles, we have $\kappa=0$ in the definition~\eqref{eq:Veffdef} of effective potential. In a spherical symmetric spacetime, these conditions lead to~\cite{Virbhadra:2008ws,Jefremov:2015gza,Jia:2017nen,Virbhadra:2022ybp}
\begin{IEEEeqnarray}{rCl}
	f(r)-\frac{r}{2}\frac{\dd f}{\dd r} &=& 0  \quad. \label{eq:frphotonsphere}
\end{IEEEeqnarray} 
Equation~\eqref{eq:frphotonsphere} provides the condition for photon sphere~\cite{Iyer:2006cn}. Substituting into the equation~\eqref{eq:sphippphi} we obtain 
\begin{IEEEeqnarray*}{rCl}
	\frac{\dd^2\bar{S}^\varphi(\varphi)}{\dd\varphi^2} &=& 0 \quad. 
\end{IEEEeqnarray*} 
Specifying the same initial conditions $\bar{S}^\varphi=1,~\frac{\dd \bar{S}^\varphi}{\dd\varphi}=S^r=0$ we can get the solution 
\begin{IEEEeqnarray}{rCl}
	\bar{S}^\varphi &=& 1 \quad, 
\end{IEEEeqnarray} 
which is unchanged when the massless particle traverses the photon sphere. As we have emphasized in last section, since $S^r$ and $\bar{S}^\varphi$ are components respect to the natural basis, our configuration $S^r=0$ and $\bar{S}^\varphi=1$ lead to a spin vector $\vec{S}$ that lies in the direction of $\vec{e}_\varphi$, and rotates with the natural basis, as we have shown in figure~\ref{fig:parallelsphere}. If two photons with identical spin are emitted from point $S$ on the photon sphere, following respectively the paths SWR and SNR, they will cancel each other out upon meeting at the receiver R. 
The same phenomenon is responsible for the absence of the glory spot in the backward scattering, as shown in section \ref{sec:unboundedorbits}. The result also confirms that there's no spin precession in Schwarzschild spacetimes~\cite{Farooqui:2013rga,Farooqui:2015ywa} for the special case of circular orbits, and here we generalize the result to arbitrary spherical symmetric spacetimes. See section \ref{sec:anglexidef} for the case of unbounded orbits.

For massive particles, we usually don't have $f-r f'/2=0$ and the solution to equations~\eqref{eq:srpphi} and~\eqref{eq:sphipphi} for initial conditions $S^r=0$ and $\bar{S}^\varphi=1$ can be written as 
\begin{IEEEeqnarray}{rCl}
	\bar{S}^\varphi &=& \cos \left(\sqrt{f-rf'/2}\varphi\right) \quad, \quad S^r= -\sqrt{f-rf'/2}\sin\left(\sqrt{f-rf'/2}\varphi\right) \quad , \label{eq:sphisrcirsolmassive}
\end{IEEEeqnarray} 
where $r$ is still the radius of the circular orbit. Therefore, if the rotating object has a radial or azimuthal component then its spin will in general not return to the original direction after a full round, contrary to the massless case. 

When the radius of the circular orbit is large, however, in an asymptotic flat spacetime we have $f(r)\to1$ and the solution~\eqref{eq:sphisrcirsolmassive} becomes 
\begin{IEEEeqnarray}{rCl}
	\bar{S}^\varphi(\varphi) &=& \cos\varphi \quad, \quad  	S^r = -\sin\varphi\quad,
\end{IEEEeqnarray} 
such that the spin vector $\vec{S}=S^r\vec{e}_r+\bar{S}^\varphi\vec{e}_\varphi$ remains a constant when the particle moves along a circular orbit, which is what one expects for motions in a flat spacetime.
In particular, for $\varphi=\pi$, we get $\bar{S}^\varphi(\pi)=-1$, and the spin vector $\vec{S}$ points in the opposite dierction of $\vec{e}_\varphi(\pi)$ but aligns with $\vec{e}_\varphi(0)$, confirming that the direction of spin vector remains unchanged.

\section{Spin precession along unbounded orbits} 
\label{sec:unboundedorbits}

For more general orbits other than the circular one, $r=r(\varphi)$ is no longer constant, making it more difficult to solve the second-order precession equation~\eqref{eq:sphippphi}, as we need to know the solution to the orbit equation~\eqref{eq:orbitequationphir} first.  

\subsection{The solution of the precession equation}
Fortunately, it turns out this won't be necessary. We can simplify the precession equation~\eqref{eq:sphippphi} by introducing a new variable $\chi$ satisfies~\cite{Dolan:2006vj}
\begin{IEEEeqnarray}{rCl}
	\frac{\dd\chi}{\dd\varphi} &=& E \left[1+\frac{L^2}{r(\varphi)^2}\right]^{-1} \quad, \label{eq:chipphi}
\end{IEEEeqnarray} 
and a new function $B[\chi(\varphi)]$ satisfies 
\begin{IEEEeqnarray}{rCl}
	\bar{S}^\varphi(\varphi) &=& \sqrt{1+\frac{L^2}{r(\varphi)^2}}B[\chi(\varphi)] \quad. \label{eq:SphiBdef}
\end{IEEEeqnarray} 
Substituting the definition~\eqref{eq:SphiBdef} of $B$ into the second order differential equation~\eqref{eq:sphippphi} of $\bar{S}^\varphi$, and changing the variable from $\varphi$ to $\chi$ according to the definition~\eqref{eq:chipphi} of $\chi$, it's directly to see that $B$ satisfies the equation 
\begin{IEEEeqnarray}{rCl}
	\frac{\dd^2B}{\dd\chi^2}+B &=& 0 \quad. 
\end{IEEEeqnarray} 
Then the precession equation can be solved once the initial conditions are given. Without loss of generality, we can assume that, at infinity, the spin aligns with the basis vector $\vec{e}_\varphi=\frac{\partial}{\partial \varphi}$, i.e., $\bar{S}^\varphi=1,~S^r=S^\theta=0$, then following the geodesic the spin will behave like 
\begin{IEEEeqnarray}{rCl}
	\bar{S}^\varphi(\varphi) &=& \sqrt{1+\frac{L^2}{r(\varphi)^2}}\cos[\chi(\varphi)]\quad,\IEEEyesnumber\IEEEyessubnumber \label{eq:barSphichisol} \\ 
	S^r(\varphi) &=& -\frac{\dd}{\dd\varphi}\bar{S}^\varphi =\frac{Er(\varphi)}{\sqrt{r(\varphi)^2+L^2}}\sin[\chi(\varphi)] + \frac{L\sqrt{E^2-V_{\text{eff}}[r(\varphi)]}}{\sqrt{r(\varphi)^2+L^2}}\cos[\chi(\varphi)]  \quad,\IEEEyessubnumber  \label{eq:Srchisol}
\end{IEEEeqnarray} 
where we have used the equation~\eqref{eq:sphipphi} and the orbit equation~\eqref{eq:orbitequationphir}.

Therefore, to get the solution of $\bar{S}^\varphi$ and $S^r$ we need to get $\chi(\varphi)$, for this we consider changing the variable $\varphi$ to $r$ in equation~\eqref{eq:chipphi}
\begin{IEEEeqnarray}{rCl}
	\frac{\dd\chi}{\dd r} &=& \frac{\dd\chi}{\dd\varphi}\frac{\dd\varphi}{\dd r} = \frac{E L}{r^2+L^2}\frac{1}{\sqrt{E^2-V_{\text{eff}}(r)}} \quad.  \label{eq:chipr}
\end{IEEEeqnarray} 
In the same way to solve the orbit equation \eqref{eq:orbitequationphir}, the differential equation~\eqref{eq:chipr} can also be integrated once to give\footnote{Note that the solution \eqref{eq:chirsolint} of $\chi(r)$ is an integral over $r$ only, and can be worked out even without knowing the orbit $r=r(\varphi)$.} 
\begin{IEEEeqnarray}{rCl}
	\chi(r) &=& \int_{r}^\infty  \frac{E L}{x^2+L^2}\frac{\dd x}{\sqrt{E^2-V_{\text{eff}}(x)}} \quad ,  \label{eq:chirsolint}
\end{IEEEeqnarray} 
and similarly, we can define $\chi_0=\chi(r_0)$, which characterizes the angle between the spins of the incoming and outgoing particles, as illustrated by figure~\ref{fig:anglephichixi}. 

 Before we continue, it's worth mentioning that although we work in the spherical symmetric spacetime for simplicity, the parallel transport equation~\eqref{eq:sphippphi} is integrable in a large class of black hole spacetimes, namely the Kerr-NUT-(A)dS ones~\cite{Connell:2008vn,Kubiznak:2008zs}, thanks to the existence of the principal conformal Killing-Yano tensor due to hidden symmetries~\cite{Frolov:2006dqt,Frolov:2008jr}. The parallel propagated orthonormal frame can be characterized by rotation angles, which are separable and integrable~\cite{Connell:2008vn,Kubiznak:2008zs}. Specifically, in a $4$d Kerr-NUT-(A)dS spacetime with spherical symmetry, only one rotation angle is needed to specify the parallel transported frame. When working in the equatorial plane, we have only $\dd r$ contributes, and the rotation angle\footnote{Cf equation (100) in~\cite{Connell:2008vn}.} in~\cite{Connell:2008vn} reduces to our $\chi$ angle given by equation \eqref{eq:chirsolint}. Therefore, the angle $\chi$ characterizes the rotation of the parallel transported frame. In the following, we are going to show how to use the solution~\ref{eq:chirsolint} of $\chi$ to characterize the precession of the spin vector, and in particular it's implication for the backward scattering.

\subsection{The angle to characterize spin precession}
\label{sec:anglexidef}

The arguments in the last subsection work for any kind of orbit, including both bounded and unbounded ones. But for the discussion of the scattering phenomena, we will focus on the unbounded orbits in the rest of this work.

As noted in section~\ref{sec:motionspin}, the particle spin is parallel transported along a geodesic, resulting in a rotation of the spin direction relative to its initial orientation. In a curved spacetime, one needs to be careful in discussing the rotation of a vector at different points. Below we will show how to use the solutions~\eqref{eq:barSphichisol} and~\eqref{eq:Srchisol} to characterize the rotation of the spin vector using $\chi$.

\begin{figure}[htpb]
	\centering
	\includegraphics[width=0.9\textwidth]{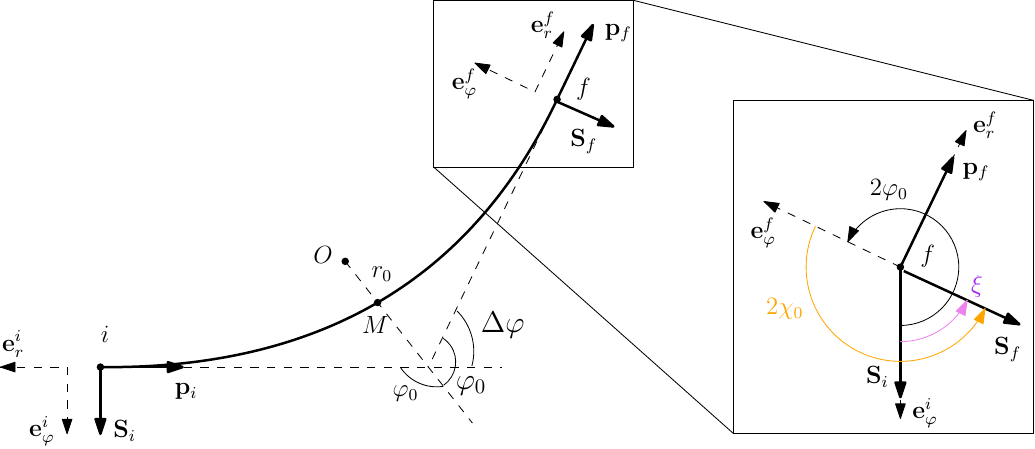}
	\caption{The illustration of spin precession along geodesics. $O$ denotes the black hole center, while $i$ and $f$ represent the initial and final point for the scattered particle (both $i$ and $f$ should be infinitely far away in practice), and $M$ is the closest point from the orbit to $O$, whose radial coordinate is  $r_0$, the smallest distance from the particle orbit to the black hole center.  $\vec{p}_i$ and $\vec{S}_i$ are the momentum and spin of the particle at the initial point $i$. $\vec{S}_i$ is defined in the particle's rest frame. While $\vec{e}_r^i$ and $\vec{e}_\varphi^i$ are unit vectors in the direction of increasing $r$ and $\varphi$ at $i$. The same applies for $\vec{p}_f,~\vec{S}_f,~\vec{e}_r^f,~\vec{e}_\varphi^f$ but respect to the final point $f$. According to the results in the main texts, we see that the angle between vectors $\vec{e}_\varphi^f$ and $\vec{e}_\varphi^i$, vectors $\vec{S}_f$ and $\vec{e}_\varphi^f$, and vectors $\vec{S}_f$ and $\vec{S}_i$ are $2\varphi_0$, $2\chi_0$ and $\xi$ respectively.}
	\label{fig:anglephichixi}
\end{figure}

Without loss of generality, we can consider an anticlockwise orbit depicted by figure~\ref{fig:anglephichixi}. We will only consider $S^r$ and $\bar{S}^\varphi$ since $S^\theta$ doesn't couple with them. Initially, the particle moving in the radial direction with speed $v$, and the components $(S^r,\bar{S}^\varphi)$ of the spin vector in the rest frame will become\footnote{Recalling that for a particle moving along $x$-axis with speed $v$, the angular momentum vector $(S^x,S^y,S^z)$ will transform to $\vec{S}$ in the rest frame of the particle 
	$\vec{S} = \left(S^x,\frac{S^y}{\sqrt{1-v^2}},\frac{S^z}{1-v^2}\right)$~\cite{Landau:1982dva}.
}
\begin{IEEEeqnarray}{rCl}
	\vec{S}_i &=& S^r\vec{e}_r^i+\frac{\bar{S}^\varphi}{\sqrt{1-v^2}}\vec{e}_\varphi^i =\frac{1}{\sqrt{1-v^2}}\vec{e}_\varphi^i  \quad, 
\end{IEEEeqnarray} 
because we have chosen the initial condition as $S^r=0$ and $\bar{S}^\varphi=1$.

Since $\chi_0=\chi(r_0)$ is the value of $\chi$ when the particle arrives at the closest point $M$, and the trajectory from $i$ to $M$ and $f$ to $M$ are symmetric, we see that at the final point $f$ one should have $\chi_f=2\chi_0$. As the final point $f$ also sits at infinity, $r_f\to\infty$, from the solutions~\eqref{eq:barSphichisol} and~\eqref{eq:Srchisol} we see that the components of the spin vector in the coordinate frame are
\begin{IEEEeqnarray}{rCl}
	\bar{S}^\varphi_f &=& \cos(2\chi_0) \quad, \quad S^r_f=E\sin(2\chi_0)=\frac{\sin(2\chi_0)}{\sqrt{1-v^2}}\quad.
\end{IEEEeqnarray} 
After transforming the components into the particle's rest frame, we find that the spin vector can be written as
\begin{IEEEeqnarray}{rCl}
	\vec{S}_f &=& \frac{\sin(2\chi_0)}{\sqrt{1-v^2}}\vec{e}_r^f+\frac{\cos(2\chi_0)}{\sqrt{1-v^2}}\vec{e}_\varphi^f \quad. 
\end{IEEEeqnarray} 
In other words, the angle between spin vector $\vec{S}_f$ in the rest frame and the unit vector $\vec{e}_\varphi^f$ in the direction of increasing $\varphi$ equals $2\chi_0$, as shown in figure~\ref{fig:anglephichixi}. 

We define the angle $\xi$ to be the angle between spin vectors $\vec{S}_i$ and $\vec{S}_f$ in the rest frame for initial and final points, in such a way that 
\begin{IEEEeqnarray}{rCl}
	\cos\xi &=& \frac{\vec{S}_f\cdot\vec{S}_i}{|\vec{S}_f||\vec{S}_i|} \quad. \label{eq:cosxidef}
\end{IEEEeqnarray} 
Note that after the scattering the particle is deflected, we have the relation 
\begin{IEEEeqnarray*}{rCl}
	\vec{e}_\varphi^f\cdot\vec{e}_\varphi^i &=& \cos(2\varphi_0)  \quad, \quad \vec{e}_r^f\cdot\vec{e}_\varphi^i=-\sin(2\varphi_0) \quad ,
\end{IEEEeqnarray*}
which is also shown in figure~\ref{fig:anglephichixi}. 
Substituting this relation into the definition~\eqref{eq:cosxidef} of $\xi$, we obtain 
\begin{IEEEeqnarray}{rCl}
	\cos\xi &=& -\sin (2\chi^0)\sin(2\varphi_0)+\cos (2\chi^0)\cos(2\varphi_0) = \cos(2\chi_0-2\phi_0) \quad. 
\end{IEEEeqnarray} 
Therefore, the angle $\xi$ can be given by $\chi_0$ and $\varphi_0$ as 
\begin{IEEEeqnarray}{rCl}
	\xi &=& 2\chi_0-2\varphi_0+2\pi \quad, \label{eq:xidef}
\end{IEEEeqnarray}
as depicted in figure~\ref{fig:anglephichixi}.

We see that $\xi$ measures the rotation of spin vectors before and after the scattering. In particular, $\cos\xi=1$ means that the direction of the spin vector doesn't change while $\cos\xi=-1$ indicates that the direction is reversed. 

\paragraph{Relativistic limit.} In general, the angle $\xi$ of a particle parallel transported along a geodesic can only be decided if the metric is explicitly given. However, the massless particle is an exception. For massless particles, we have $U^\mu U_\mu=0$, then the orthogonal relation~\eqref{eq:smuUmuorthogonal} indicates that in our case the spin $3-$vector is orthogonal to the $3$-momentum,  and the photon spin will always be reversed ($\chi_0=\pi/2$) in the backward scattering, resulting in $\cos\xi=-1$ \cite{Dolan:2006vj,Crispino:2009xt}. This is valid for any spherical symmetric spacetime, including Schwarzschild and \acrshort{rn} ones, as depicted in figures \ref{fig:cosxivarphi1357piq0} and \ref{fig:cosxivarphi1357pih0p9}. 

 In fact, the rotation of the spin vector compared to the initial condition is due to the parallel transportation of our reference frame. Following \cite{Farooqui:2013rga,Farooqui:2015ywa} one can see that the normal vectors $\vec{e}_r$ and $\vec{e}_{\varphi}$ constitute an orthonormal basis lies in the equatorial plane. The orthonormal relation~\eqref{eq:smuUmuorthogonal} implies that the spin vector will rotate along the reference frame, in other words, measured in the basis spanned by $\vec{e}_r$ and $\vec{e}_{\varphi}$, there will be no precession of the spin vector, in consistent with the results in~\cite{Farooqui:2013rga,Farooqui:2015ywa,Lusk:2024fpo}\footnote{See corollary 7.2 in \cite{Farooqui:2015ywa}, for example}. As in the case of circular orbits (mentioned in section \ref{sec:circularorbits}), our results valid for a general spherical symmetric spacetime, not restrict to the Schwarzschild one.

We emphasized again that in the scenario of backward scattering, $\cos\xi=-1$ indicates that the direction of spin is reversed compared to its initial direction. As is illustrated by figure~\ref{fig:parallelunbounded}, where the spin vector $1$ becomes $1'$ after the backward scattering following the path SBMCS. On the other hand, following the path SAMDS perpendicular to the equatorial plane instead, 
the direction of the spin vector $1$ remains unchanged, which becomes $1''$ after the backward scattering. Since $1'$ and $1''$ are opposite to each other, they will cancel perfectly, and there would be no glory spot for the scattering of massless particles with non-vanishing spin~\cite{Crispino:2009xt}, in contradiction to the WKB approximation for spinless particles. 

\section{Examples: Schwarzschild and \acrlong{rn} spacetimes}
\label{sec:examplesSchRN}

Although the relativistic behaviour of the spin doesn't depend on the metric, to fully analyse the precession of massive particles, we have to explicitly integrate \eqref{eq:chirsolint} based on given metrics. This section will consider two spherical symmetric spacetimes, the Schwarzschild and \acrshort{rn} ones. 

\subsection{Schwarzschild spacetime}
\label{sec:exampleSch}
For Schwarzschild spacetime, the metric reads 
\begin{IEEEeqnarray}{rCl}
	f(r) &=& 1-\frac{2M}{r} \quad. 
\end{IEEEeqnarray} 
Introducing dimensionless variables~\cite{Pang:2018jpm}\footnote{In the following we will see that the definition of $\eta$ allows us to discuss the non-relativistic limit more easily.} 
\begin{IEEEeqnarray}{rCl}
	u &=& \frac{M}{r} \quad, \quad \eta=\frac{bv}{M} \quad, \label{eq:uetadef}
\end{IEEEeqnarray} 
we can write $\varphi_0$ and $\chi_0$ as (by setting $r=r_0$ and changing variable to $u$ in equations \eqref{eq:varphirsolint} and \eqref{eq:chirsolint}) 
\begin{IEEEeqnarray}{rCl}
	\varphi_0 &=& \int_{0}^{u_2}\frac{\dd u}{\sqrt{2}\sqrt{P_s(u)}}  \quad,\label{eq:varphi0Schint} \\ 
	\chi_0 &=& \int_0^{{u_2}}\frac{\sqrt{1-v^2}}{\sqrt{2}(\eta^2u^2+1-v^2)}\frac{\dd u}{\sqrt{P_s(u)}} \quad,  \label{eq:chi0Schint}
\end{IEEEeqnarray} 
where $P_S(u)$ is a cubic polynomial of $u$~\cite{Liu:2015zou}
\begin{IEEEeqnarray}{rCl}
	P_S(u) &=& u^3-\frac{1}{2}u^2+\frac{1-v^2}{\eta^2}u+\frac{v^2}{2\eta^2}=(u_3-u)(u_2-u)(u-u_1) \quad, \label{eq:upolySch}
\end{IEEEeqnarray} 
and $u_1,~u_2,~u_3$ are three roots of the polynomial $P_S(u)$. In the scattering scenario, we typically find $u_1\leq 0\leq u<u_2\leq u_3$, so $u_2$ is the smallest positive root, whose inverse is the closest distance from the orbit to the black hole center, $u_2=\frac{M}{r_0}$. 

The two integrals~\eqref{eq:varphi0Schint} and~\eqref{eq:chi0Schint} for $\varphi_0$ and $\chi_0$ can be evaluated to elliptic functions~\cite{Gradshteyn:1996table}
\begin{IEEEeqnarray}{rCl}
	\varphi_0 &=& \frac{\sqrt{2}\mathrm{F}\left(\phi_s,q_s\right),}{\sqrt{u_3-u_1}} \quad, \label{eq:varphi0Sch}\\ 
	\chi_0 &=& \frac{\sqrt{1-v^2}}{\eta^2u_3^2+1-v^2} \varphi_0 +\frac{\ii \eta (u_2-u_3)}{\sqrt{2}\sqrt{u_3-u_1}}\left[\frac{1}{\eta_{2}\eta_{3}}\mathrm{\Pi}\left(q_s\frac{\eta_{3}}{\eta_{2}},\phi_s,q_s\right)\right.\nonumber \\ 
		   &&-\left.\frac{1}{\eta_{2}^*\eta_{2}^*}\mathrm{\Pi}\left(q_s\frac{\eta_{3}^*}{\eta_{2}^*},\phi_s,q_s\right)\right]\quad,   \label{eq:chi0Sch}
\end{IEEEeqnarray} 
where ${}^*$ denotes the complex conjugation and 
\begin{IEEEeqnarray}{rCl}
	\phi_s &=& \sin^{-1}\left(\sqrt{\frac{u_2(u_3-u_1)}{u_3(u_2-u_1)}}\right) \quad, \quad q_s = \frac{u_2-u_1}{u_3-u_1}\quad, \\ 
	\eta_{2}&=& \eta u_2-\ii\sqrt{1-v^2}\quad,\quad \eta_{3}=\eta u_3-\ii\sqrt{1-v^2}\quad. \label{eq:eta2eta3def}
\end{IEEEeqnarray} 
The subscripts $s$ indicate that these quantities are defined for Schwarzschild spacetime. $\mathrm{F}(\phi,m)$ and $\mathrm{\Pi}(n,\phi,m)$ are incomplete elliptic integrals of the first and third kinds, respectively~\cite{Gradshteyn:1996table}, whose definitions are listed in appendix~\ref{sec:ellipticfunction} for convenience.

Noting the fact that $\mathrm{\Pi}(z^*,\phi,m)=\mathrm{\Pi}(z,\phi,m)^*$, the two terms in the last bracket of the expression~\eqref{eq:chi0Sch} are complex conjugate to each other. Therefore, $\chi_0$ is real as it should be, despite the appearance of the complex numbers in the expression.

\begin{figure}[htpb]
	\centering
	\includegraphics[width=0.6\textwidth]{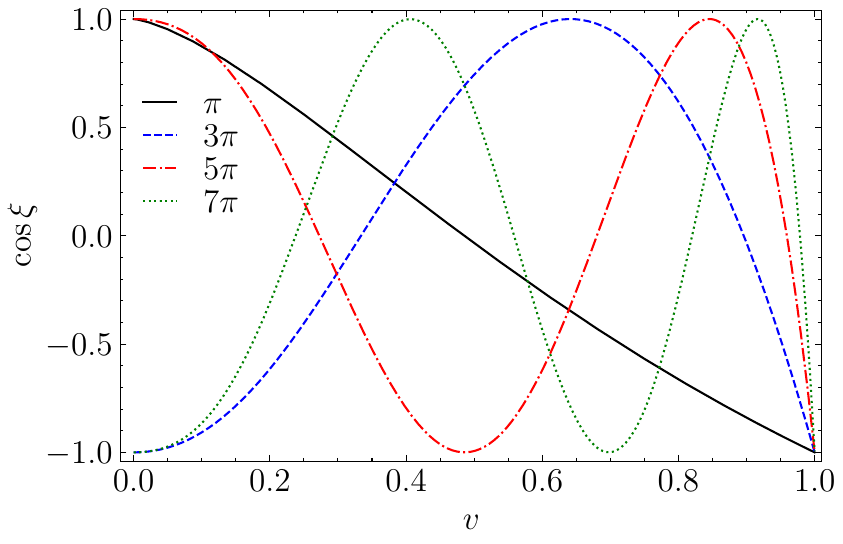}
	\caption{The behaviour of $\cos\xi$ for different $\Delta\varphi$ in the backward scattering, Schwarzschild case. The black solid line, blue dashed line, red dash-dotted line and green dotted line represent the behaviour of $\cos\xi$ for $\Delta\varphi=\pi,~3\pi,~5\pi$ and $7\pi$, respectively. We see that for massless particles $v=1$ we always have $\cos\xi=-1$, which indicates the vanishing of glory spot in the backward scattering. For $v=0$, on the other hand, according to the definition~\eqref{eq:deltaphidef} of deflection angle $\Delta\varphi$ we see that $\varphi_0=\pi,~2\pi,~3\pi$ and $4\pi$, then from equation~\eqref{eq:cosxiv0Sch} we see that $\cos\xi=-\cos\varphi_0=1$ for $\varphi_0=\pi$ and $3\pi$ while $\cos\xi=-1$ for $\varphi_0=2\pi$ and $4\pi$, as we can expect.}
	\label{fig:cosxivarphi1357piq0}
\end{figure}

In figure~\ref{fig:cosxivarphi1357piq0}, we plot the behaviour of $\cos\xi$ for different deflection angles $\Delta \varphi$ during backward scattering in the Schwarzschild case. We find that for $v=1$ one always gets $\cos\xi=-1$, and there will be perfect cancellation of spins in the backward scattering, for particles coming from perpendicular orbits, as we have mentioned in the end of section \ref{sec:unboundedorbits}. On the other hand, for non-relativistic particles $v\approx0$, $\cos\xi$ is no longer a constant.

\paragraph{Non-relativistic limit.} To see how the spin precesses in the non-relativistic limit for backward scattering, we can get a simpler expression of $\chi_0$ by expansion. 
In fact, for the case under consideration, the lensing of the particle would be strong in the sense that $\delta^2=\eta-\eta_c$ is small (we have introduced $\delta^2$ instead of $\delta$ for convenience, see below) \cite{Bozza:2002zj,Liu:2015zou,Pang:2018jpm}. Substituting $v=0$ into the polynomial~\eqref{eq:upolySch} we see that the first root $u_1$ vanishes, then $\chi_0$ becomes 
\begin{IEEEeqnarray}{rCl}
	\chi_0 &=& \frac{\varphi_0}{\eta^2u_3^2-1}+ \frac{\ii\eta(u_2-u_3)}{\sqrt{2u_3}}\left[\frac{1}{(\eta u_2-\ii)(\eta u_3-\ii)}\mathrm{\Pi}\left(\frac{u_2(\eta u_3-\ii)}{u_3(\eta u_2-\ii)},\frac{u_2}{u_3}\right)\right. \nonumber \\ 
		   &&\left.+\frac{1}{(\eta i_2+\ii)(\eta u_3+\ii)}\mathrm{\Pi}\left(\frac{u_2(\eta u_3+\ii)}{u_3(\eta u_2+\ii)},\frac{u_2}{u_3}\right)\right] \quad, \label{eq:chi0Schexpmed}
\end{IEEEeqnarray} 
where $\Pi(n,m)$ represents the complete elliptic integral of the third kind.
For Schwarzschild spacetime, we have $\eta_c=4$ and when $\eta=\eta_c+\delta^2$ for small $\delta$ (see appendix~\ref{sec:rootexpansion}), we have 
\begin{IEEEeqnarray}{rCl}
	u_2 &=& u_c-\frac{1}{4\sqrt{2}}\delta \quad, \quad u_3= u_c+\frac{1}{4\sqrt{2}}\delta \quad,
\end{IEEEeqnarray} 
for $u_c=1/4$ corresponds to the position of particle sphere~\cite{Pang:2018jpm}, such that the deflection angle $\Delta\varphi\to\infty$ when $u_2\to u_c$.

Substituting these expansions into equation \eqref{eq:chi0Schexpmed}, to the leading order of $\delta$ we obtain
\begin{IEEEeqnarray*}{rCl}
	\chi_0 &=& \frac{1}{2}\varphi_0+ \delta\left[\mathrm{\Pi}\left(1-\frac{1+\ii}{2}(\sqrt{2}\delta),1-\sqrt{2}\delta\right)-\mathrm{\Pi}\left(1-\frac{1-\ii}{2}(\sqrt{2}\delta),1-\sqrt{2}\delta\right)\right]\quad. 
\end{IEEEeqnarray*} 
The elliptic functions in the bracket can be simplified using equation~\eqref{eq:PiPistarsum}. Substituting $z=\frac{1+\ii}{2}$ into equation~\eqref{eq:PiPistarsum} and replacing $\delta$ by $\sqrt{2}\delta$ we obtain 
\begin{IEEEeqnarray*}{rCl}
	&&\mathrm{\Pi}\left(1-\frac{1+\ii}{2}(\sqrt{2}\delta),1-\sqrt{2}\delta\right)-\mathrm{\Pi}\left(1-\frac{1-\ii}{2}(\sqrt{2}\delta),1-\sqrt{2}\delta\right) \\
	 =&& \frac{\sqrt{2}}{\delta}\mathrm{Re}\left(-\sqrt{2}\ii\cosh^{-1}\left(\sqrt{\frac{1+\ii}{2}}\right)\right) =\frac{\pi}{2\delta}  \quad.
\end{IEEEeqnarray*} 
This results in 
\begin{IEEEeqnarray}{rCl}
	\chi_0 &=& \frac{\varphi_0}{2}+\frac{\pi}{2} \quad. \label{eq:chi0Schexp}
\end{IEEEeqnarray} 
Substituting the value of $\chi_0$ into the definition~\eqref{eq:xidef} of angle $\xi$, we have
\begin{IEEEeqnarray}{rCl}
	\cos\xi &=& \cos(2\chi_0-2\varphi_0+2\pi)=\cos(-\varphi_0+\pi)=-\cos(\varphi_0) \quad , \label{eq:cosxiv0Sch}
\end{IEEEeqnarray} 
which is consistent with figure~\ref{fig:cosxivarphi1357piq0}. Since $\cos\xi=-1$ would mean a perfect cancellation in the backward scattering, for non-relativistic particles with $v\approx 0$, the cancellation will occur for $\varphi_0=2n\pi$ (deflection angle $\Delta\varphi=(2n-1)\pi$) with integer $n\geq 1$. However, unlike the case of massless particles, where we always have $\cos\xi=-1$ in the backward direction, for non-relativistic particles we have $\cos\xi=1$ for $\varphi_0=(2n-1)\pi$ for $n\geq 1$. Since the backward scattering is dominated by the one with $\varphi_0=\pi$, 
the glory spot will not vanish for the non-relativistic particles.

\subsection{\acrlong{rn} spacetime}
In this section, we will further set $4\pi\varepsilon_0=1$ for the vacuum permittivity $\varepsilon_0$. Then the metric of \acrshort{rn} spacetime reads
\begin{IEEEeqnarray}{rCl}
	f(r) &=& 1-\frac{2M}{r}+\frac{Q^2}{r^2} \quad. 
\end{IEEEeqnarray} 
In addition to $u$ and $\eta$ defined in~\eqref{eq:uetadef}, we also introduce another dimensionless parameter 
\begin{IEEEeqnarray}{rCl}
	h &=& \frac{Q}{M} \quad, 
\end{IEEEeqnarray} 
and $\varphi_0$ and $\chi_0$ in \acrshort{rn} spacetime can be written as 
\begin{IEEEeqnarray}{rCl}
	\varphi_0 &=& \int_0^{u_2}\frac{\dd u}{h\sqrt{P_{RN}(u)}} \quad, \\  \label{eq:varphi0RNint}
	\chi_0 &=& \int_0^{{u_2}}\frac{\sqrt{1-v^2}}{(\eta^2u^2+1-v^2)}\frac{\dd u}{h\sqrt{P_{RN}(u)}} \quad,  \label{eq:chi0RNint}
\end{IEEEeqnarray} 
where $P_{RN}(u)$ is a quartic polynomial of $u$ 
\begin{IEEEeqnarray}{rCl}
	P_{RN}(u) &=& -u^4+\frac{2}{h^2}u^3-\left(-\frac{1}{h^2}+\frac{1-v^2}{\eta^2}\right)u^2+\frac{2(1-v^2)}{\eta^2h^2}u+\frac{v^2}{\eta^2h^2}\nonumber \\ 
			  &=&  (u_4-u)(u_3-u)(u_2-u)(u-u_1)  \quad, \label{eq:upolyRN}
\end{IEEEeqnarray} 
and $u_1\leq0<u_2 \leq u_3<u_4$ are four roots of the polynomial $P_{RN}(u)$. Similar to the Schwarzschild case, for unbounded orbits $u$ is confined between $0$ and the first positive root $u_2$, $0\leq u\leq u_2$. In \acrshort{rn} case, the integral in~\eqref{eq:varphi0RNint} and~\eqref{eq:chi0RNint} can also be worked out as elliptic functions
\begin{IEEEeqnarray}{rCl}
	\varphi_0 &=& \frac{2\mathrm{F}\left(\phi_{rn},q_{rn}\right),}{h\sqrt{(u_3-u_1)(u_4-u_2)}} \quad, \label{eq:varphi0RN}\\ 
	\chi_0 &=& \frac{\sqrt{1-v^2}}{\eta^2u_3^2+1-v^2} \varphi_0 +\frac{\ii \eta (u_2-u_3)}{h\sqrt{(u_3-u_1)(u_4-u_2)}}\left[\frac{1}{\eta_{2}\eta_{3}}\mathrm{\Pi}\left(p_{rn}^2\frac{\eta_{3}}{\eta_{2}},\phi_{rn},q_{rn}\right)\right.\nonumber \\ 
		   &&-\left.\frac{1}{\eta_{2}^*\eta_{2}^*}\mathrm{\Pi}\left(p_{rn}^2\frac{\eta_{3}^*}{\eta_{2}^*},\phi_{rn},q_{rn}\right)\right]\quad,   \label{eq:chi0RN}
\end{IEEEeqnarray} 
where we have defined 
\begin{IEEEeqnarray}{rCl}
	\phi_{rn} &=& \sin^{-1}\left(\sqrt{\frac{u_2(u_3-u_1)}{u_3(u_2-u_1)}}\right) \quad, \quad q_{rn} = \frac{(u_2-u_1)(u_4-u_3)}{(u_3-u_1)(u_4-u_2)}\quad, \quad p_{rn}= \sqrt{\frac{u_2-u_1}{u_3-u_1}} \quad,\nonumber \\ 
\end{IEEEeqnarray} 
and the subscripts $rn$ denotes that these quantities are defined for \acrshort{rn} spacetime.
$\eta_2,~\eta_3$ are still given by equation~\eqref{eq:eta2eta3def}, just now $u_2$ and $u_3$ should be the corresponding roots of the quartic equation~\eqref{eq:upolyRN} instead.

\begin{figure}[htpb]
	\centering
	\includegraphics[width=0.6\textwidth]{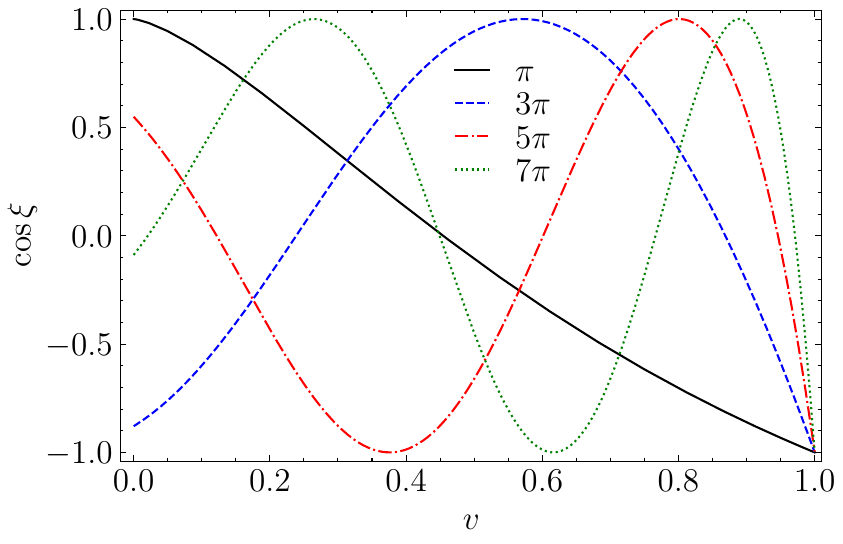}
	\caption{The behaviour of $\cos\xi$ for different $\Delta\varphi$ in the backward scattering, \acrshort{rn} case with $h=Q/M$, for black hole charge $Q$. As in figure~\ref{fig:cosxivarphi1357piq0}, the black solid line, blue dashed line, red dash-dotted line and green dotted line represent the behaviour of $\cos\xi$ for $\Delta\varphi=\pi,~3\pi,~5\pi$ and $7\pi$, respectively. Again we see that for $v=1$ we have $\cos\xi=-1$, which is independent of the charge $Q$. While for $v=0$, $\cos\xi$ becomes a charge-dependent quantity, as we can see from equation~\eqref{eq:chi0RNexp}.}
	\label{fig:cosxivarphi1357pih0p9}
\end{figure}

Figure~\ref{fig:cosxivarphi1357pih0p9} illustrates the behaviour of $\cos\xi$ for different deflection angles $\Delta \varphi$ in the backward scattering for \acrshort{rn} spacetime with $h=Q/M=0.9$. Again, we see that for $v=1$ we always have $\cos\xi=-1$ as we have expected. The same cancellation for spinning particles in the backward scattering will occur as well as in the Schwarzschild spacetime. 

In contrast to the Schwarzschild case, however, in \acrshort{rn} spacetime $\cos\xi$ for $v=0$ will depend on the black hole charge $h=Q/M$, in a way such that different deflection angles $\Delta\varphi$ lead to different values of $\cos\xi$, except for deflection angle $\Delta\varphi=\pi$. 

\paragraph{Non-relativistic limit.} 

Similar to the Schwarzschild case, we consider the strong lensing expansion in \acrshort{rn} spacetime
\begin{IEEEeqnarray}{rCl}
	\eta &=&\eta_c+\delta^2	  \quad, \quad u_2=u_c-\sqrt{\alpha}\delta \quad,\quad u_3=u_c+\sqrt{\alpha}\delta \quad , \label{eq:etacu2u3RN}
\end{IEEEeqnarray} 
Note that $\eta_c$ is the critical value of $\eta$ in \acrshort{rn} spacetime, whose expression is given by equation~\eqref{eq:etacrn} in appendix~\ref{sec:rootexpansion}, $u_c$ is the double root of equation~\eqref{eq:uceqnoeta}, and $\alpha$ is given by~\eqref{eq:alphaRNv0}. 
Substituting the expansion~\eqref{eq:etacu2u3RN} into equation~\eqref{eq:chi0RN} and also note in the non-relativistic case $u_1=0$, to the first order we can expand $\chi_0$ as 
\begin{IEEEeqnarray}{rCl}
	\chi_0 &=& \frac{\varphi_0}{1+\eta_c^2u_c^2} + \frac{2\ii \eta_c\sqrt{\alpha}\delta}{h\sqrt{(u_4-u_c)u_c}}\left[\frac{1}{\left(-\ii+\eta_c u_c\right)^2}\mathrm{\Pi}\left(1-\frac{2\ii\sqrt{\alpha}\delta}{u_c(\ii-\eta_cu_c)},1-\frac{2u_4\sqrt{\alpha}\delta}{(u_4-u_c)u_c}\right)\right.\nonumber \\ 
		   &&+\left.\frac{1}{\left(\ii+\eta_c u_c\right)^2}\mathrm{\Pi}\left(1-\frac{2\ii\sqrt{\alpha}\delta}{u_c(\ii+\eta_cu_c)},1-\frac{2u_4\sqrt{\alpha}\delta}{(u_4-u_c)u_c}\right)\right] \quad. 
\end{IEEEeqnarray} 
Substituting equation~\eqref{eq:alphaRNv0} in last equation, taking the limit $\delta\to0$ and using equation~\eqref{eq:Picosh}, after some algebraical simplification, we obtain
\begin{IEEEeqnarray}{rCl}
	\chi_0 &=& \frac{\varphi_{0}}{1+\eta_{c}^2u_{c}^2} +\frac{\eta_c}{h}\left[\frac{\sech^{-1}\left(\sqrt{\frac{u_4(1-\ii \eta_{c})}{u_4-u_{c}}}\right)}{\sqrt{1-\ii\eta_{c}}(-1+\ii\eta_{c}u_{c})}-\frac{\sech^{-1}\left(\sqrt{\frac{u_4(1+\ii \eta_{c})}{u_4-u_{c}}}\right)}{\sqrt{1+\ii\eta_{c}}(1+\ii\eta_{c}u_{c})}\right] \quad, \label{eq:chi0RNexp}
\end{IEEEeqnarray} 
where $\sech^{-1}(z)$ is the inverse of the hyperbolic secant $\sech(z)=\frac{1}{\cosh(z)}$. And $\eta_c$ is given by equation~\eqref{eq:etacrn}, while $u_c$ and $u_4$ are the double root and the largest root of the quartic equation~\eqref{eq:uceqnoeta}, respectively. Setting $\eta_c=4,~u_c=\frac{1}{4}$ and $u_4=\frac{2}{h^2}$ then taking the $h\to0$ limit we recover the Schwarzschild result~\eqref{eq:chi0Schexp}.

\begin{figure}[htpb]
	\centering
	\includegraphics[width=0.6\textwidth]{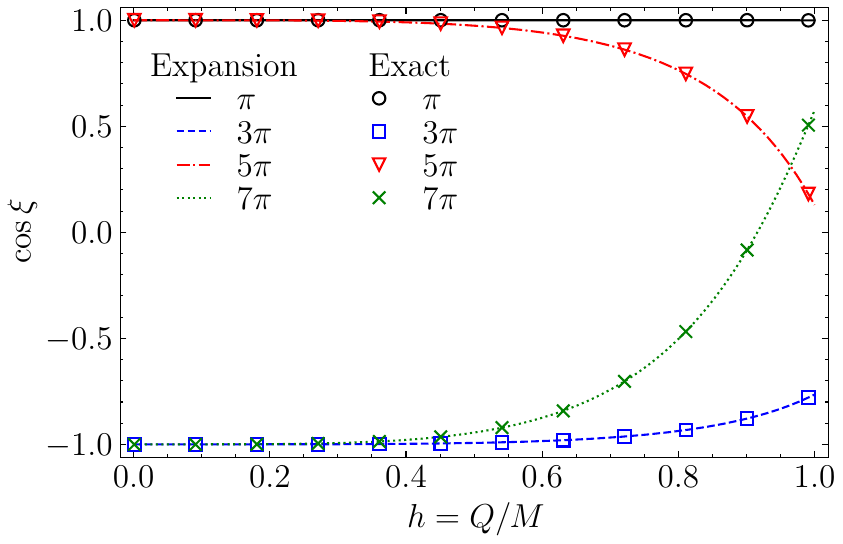}
	\caption{The comparison between the cosine $\cos\xi$ for the exact value~\eqref{eq:chi0RN} and the expansion~\eqref{eq:chi0RNexp} of $\chi_0$. The continuous lines have the same meaning as in figure~\ref{fig:cosxivarphi1357piq0}. While the black circle, blue square, red triangle and blue cross mark represent the value of $\cos\xi$ for $\chi_0$ from the exact expression~\eqref{eq:chi0RN}.}
	\label{fig:cosxivarphi1357pihcomp}
\end{figure}

In figure~\ref{fig:cosxivarphi1357pihcomp} we compared exact result \eqref{eq:chi0RN} and our expansion~\eqref{eq:chi0RNexp} in the \acrshort{rn} spacetime. We see that the expansion is pretty accurate compared to the exact value, which shows the validity of our expansion \eqref{eq:chi0RNexp} in the strong field limit.

\section{Discussion}
\label{sec:summary}
In this paper, we have considered the spin precession of both massless and massive particles in spherical symmetric spacetimes. The spin is parallel transported along the geodesics, 
and for massless particles, this precession accounts for the absence of the glory spot in the backward scattering, a phenomenon that can't be explained at the leading order of the WKB approximation.

To see how the spin precesses as the particle moves along the geodesic, we derived the precession equation governing the components of the spin vector respective to the natural basis. For circular orbits, the radial coordinate $r$ remains a constant, and the precession equation is quite easy to solve. When $v=1$, the circular orbits can only live on the photon sphere, and the particle spin rotates along with the natural basis, indicating that two photons emitted from the same point will cancel each other out upon meeting; while for massive particles, the particle spin also rotates in general but with different paces depending on the radius. For a very large radius, the direction of the spin will remain unchanged if the spacetime is asymptotically flat. 

The precession equation for unbounded orbits is harder to solve because now the radial coordinate $r$ is no longer a constant. Yet the equation can be greatly simplified by introducing the new variable $\chi$ satisfies the equation \eqref{eq:chipphi}. At infinities, $\chi$ characterizes the angle between spin vector $\vec{S}$ and $\vec{e}_\varphi$, the unit vector in the direction of increasing azimuth coordinate $\varphi$. The spin precession can be measured by the angle $\xi$ between spins before and after the scattering, which depends on $\chi$ and the deflection angle. Then we have shown that for massless particles one always has $\cos\xi=-1$ in the backward scattering, indicating a perfect cancellation and the vanishing of the glory spot. This result is a direct consequence of the orthogonal relation \eqref{eq:smuUmuorthogonal} between the spin $S^\mu$ and the $4$-velocity $U^\mu$ of a particle, and doesn't depend on the explicit form of the metric. 

To get the spin precession of massive particles, however, one has to work with explicit metrics. We have considered two cases, the Schwarzschild and \acrshort{rn} spacetimes. For both cases, we can solve both the geodesic and precession equations explicitly using elliptic functions. And $\cos\xi$ approaches $-1$ for $v$ approaches the speed of light, 
 in agreement with our discussion of massless particles. The non-relativistic limit with $v\approx 0$ is also considered. And we have derived the expansion based on strong lensing approximation. The result \eqref{eq:chi0Schexp} is very simple in Schwarzschild spacetime, while the corresponding one \eqref{eq:chi0RNexp} in \acrshort{rn} spacetime is still a bit complicated. Both results agree with the exact result for $v=0$ with high accuracy and can be applied to the scattering of slow particles.

\medskip

In this work, we have focused on the geodesic equations rather than the \acrshort{mpeq} equation by ignoring the spin-orbit coupling, and as a consequence, there is no back action from the spin to the particle trajectory. A full analysis for the spin precession based on the \acrshort{mpeq} for both massless and massive particles is needed in future works, to check whether the glory spot still vanishes in the backward scattering for massless particles with spin. On the other hand, even without the discussion of glory scattering, the precession of spins in curved spacetime is interesting on its own, and we can expand the discussion to non-spherical symmetric spacetimes, like the axisymmetric ones including Kerr spacetime for its observational relevance, and the class of Kerr-NUT-(A)dS spacetimes where the parallel transport equations are separable and integrable. More realistically, we can go beyond the deflection and precession of a single particle and consider a congruence of particle rays~\cite{Li:2024oke}, then investigate the behaviour of the spinning particles based on the geodesic deviation equation. 

\acknowledgments

The authors appreciate the financial support from The Central Guidance on Local Science and Technology Development Fund of Sichuan Province (2024ZYD0075). GD is also supported by the Fundamental Research Founds of China West Normal University (No.22kA005). YX is supported by the Doctoral Initiation Fund of West China Normal University (22kE040), the Open Fund of Key Laboratory of Astroparticle Physics of Yunnan Province (2022Zibian3), and the National Natural Science Foundation of China (NSFC, Grant No. 12303048).

\appendix 
\section{The parallel transport of Dirac fermion and the corresponding spin vector}
\label{sec:paralleldiracspinor}

To be able to handle fermions in \acrshort{gr}, we need to use tetrad formalism~\cite{Weinberg:1972kfs,Poplawski:2009fb}. In fact, for a Dirac fermion that is parallel transported along a geodesic parametrized by $\tau$, we have  
\begin{IEEEeqnarray}{rCl}
	\frac{\mathrm{D}\psi}{\dd\tau} &=& \frac{\dd\psi}{\dd \tau}-\Gamma_\mu\psi \frac{\dd x^\mu}{\dd\tau} =0   \quad, \label{eq:DpsiGammamu}
\end{IEEEeqnarray} 
where the spin connection is defined by\footnote{As one can verify, under a Lorentz transformation $U$, $\Gamma_\mu$ transforms to $\Gamma_\mu\to (\partial_\mu U)U^{-1}+U\Gamma_\mu U^{-1}$, and the covariant differential $\mathrm{D}\psi=\dd\psi-\Gamma_\mu\psi\dd x^\mu$ transforms like a Dirac spinor as it should be.}~\cite{Poplawski:2009fb}
\begin{IEEEeqnarray}{rCl}
	\Gamma_\mu &=&-\frac{\ii}{2}\Gamma_{abc}\Sigma^{ab}e^c_{~\mu}  \quad. 
\end{IEEEeqnarray} 
The tetrad fields $e_c^{~\mu}$ satisfy $\eta^{ab}e_a^{~\mu}e_b^{~\nu}=g^{\mu\nu}$ with Minkowski metric $\eta^{ab}$\footnote{We will use $a,~b,~c,\cdots$ to denote the Lorentz index while using $\mu,~\nu,~\lambda,\cdots$ to denote the coordinate index.}. The \emph{Ricci rotation coefficients} $\Gamma_{abc}$~\cite{Landau:1982dva} and the generator $\Sigma^{ab}$ for Lorentz group in the Dirac fermion representation read
\begin{IEEEeqnarray*}{rCl}
	\Gamma_{abc} &=& (e_{a\mu})_{;\nu}e_b^{~\mu}e_c^{~\nu}  \quad,  \\ 
	\Sigma^{ab} &=& \frac{\ii}{4}[\gamma^a,\gamma^b] \quad,  
\end{IEEEeqnarray*} 
where $\gamma^a$ are the usual $\gamma$-matrices and $\displaystyle (e_{a\mu})_{;\nu}=\partial_\nu e_{a\mu}-\Gamma^\lambda_{\mu\nu}e_{a\lambda}$. 

Note that $\bar{\psi}\psi$ is a Lorentz scalar, we have $\displaystyle\frac{\mathrm{D}(\bar{\psi}\psi)}{\dd\tau}=\frac{\dd(\bar{\psi}\psi)}{\dd\tau}$. Therefore, the corresponding parallel transport of $\bar{\psi}$ reads
\begin{IEEEeqnarray}{rCl}
	\frac{\mathrm{D}\bar{\psi}}{\dd\tau} &=& \frac{\dd\bar{\psi}}{\dd \tau}+\bar{\psi}\Gamma_\mu \frac{\dd x^\mu}{\dd\tau} =0   \quad.  \label{eq:DbarpsiGammamu}
\end{IEEEeqnarray} 
Furthermore, the tetrads $e_a^{~\mu}$ are auto-parallel transported along a geodesic, i.e.,
\begin{IEEEeqnarray*}{rCl}
	\frac{\mathrm{D}e_a^{~\mu}}{\dd \tau} &=& \frac{\dd}{\dd\tau}e_a^{~\mu}+\Gamma^\mu_{\nu\lambda}e_a^{~\nu}\frac{\dd x^\lambda}{\dd\tau}-\Gamma^b_{ac}e_b^{~\mu}e^c_{~\nu}\frac{\dd x^\nu}{\dd\tau}=0  \quad. 
\end{IEEEeqnarray*} 
Then for the position dependent $\gamma$-matrices $\gamma^\mu=e_a^{~\mu}\gamma^a$ we have 
\begin{IEEEeqnarray}{rCl}
	\frac{\dd \gamma^\mu}{\dd\tau}+\Gamma^\mu_{\nu\lambda}\gamma^\nu\frac{\dd x^\lambda}{\dd\tau}&=&\Gamma_{abc}e^{b\mu}e^c_{~\nu}\gamma^a\frac{\dd x^\nu}{\dd \tau}  \quad. \label{eq:Gammabacgammaa}
\end{IEEEeqnarray} 

Now we are ready to show that the spin vector $S^\mu=\bar{\psi}\gamma^\mu\gamma_5\psi$ will be parallel transported along a geodesic if the Dirac fermion $\psi$ does. In fact, by substituting equations~\eqref{eq:DpsiGammamu},~\eqref{eq:DbarpsiGammamu} and~\eqref{eq:Gammabacgammaa}, we obtain 
\begin{IEEEeqnarray}{rCl}
	\frac{\mathrm{D}SS^\mu}{\dd\tau} &=& \frac{\dd S^\mu}{\dd\tau}+\Gamma^\mu_{\nu\lambda}S^{\nu} \frac{\dd x^\lambda}{\dd\tau} \quad,  \nonumber \\ 
									&=&  \frac{\dd \bar{\psi}}{\dd\tau} \gamma^\mu\gamma_5\psi +\bar{\psi}\gamma^\mu\gamma_5 \frac{\dd\psi}{\dd\tau} + \bar{\psi}\left(\frac{\dd\gamma^\mu}{\dd\tau}+\Gamma^\mu_{\nu\lambda}\gamma^\nu \frac{\dd x^\lambda}{\dd\tau}\right)\gamma_5\psi \quad, \nonumber  \\ 
									&=& -\bar{\psi}\Gamma_\nu\gamma^\mu\gamma_5\psi \frac{\dd x^\nu}{\dd\tau} +\bar{\psi}\gamma^\mu\gamma_5\Gamma_\nu\psi \frac{\dd x^\nu}{\dd\tau}+ \bar{\psi}\Gamma_{abc}e^{b\mu}e^c_{~\nu}\gamma^a\gamma_5\psi\frac{\dd x^\nu}{\dd \tau} \quad, \nonumber \\ 
	 &=& \bar{\psi}\left(-\Gamma_\nu\gamma^\mu+\gamma^\mu\Gamma_\nu\right)\gamma_5\psi \frac{\dd x^\nu}{\dd\tau} + \bar{\psi}\Gamma_{abc}e^{b\mu}e^c_{~\nu}\gamma^a\gamma_5\psi\frac{\dd x^\nu}{\dd \tau}  \quad, \label{eq:DsmudtauGamma}
\end{IEEEeqnarray} 
where we have used $\gamma_5\Gamma_\nu=\Gamma_\nu\gamma_5$ and equation~\eqref{eq:Gammabacgammaa}. To further simplify the result, we recall the result in Dirac algebra 
\begin{IEEEeqnarray*}{rCl}
	\left[\Sigma^{ab},\gamma^d\right]  &=&  \ii\left(\eta^{bd}\gamma^a-\eta^{ad}\gamma^b\right)\quad, 
\end{IEEEeqnarray*} 
then we have 
\begin{IEEEeqnarray*}{rCl}
	-\Gamma_\nu \gamma^\mu +\gamma^\mu\Gamma_\nu&=& \frac{\ii}{2}e^c_{~\nu}\Gamma_{abc}e_d^{~\mu}\left[\Sigma^{ab},\gamma^d\right]  \quad,  \\ 
												&=& \frac{\ii}{2}e^c_{~\nu}e_d^{~\mu}\Gamma_{abc}(\ii)\left(\eta^{bd}\gamma^a-\eta^{ad}\gamma^b\right) \quad,   \\ 
												&=& \frac{1}{2}e^c_{~\nu}e^{a\mu}\gamma^b\left(\Gamma_{abc}-\Gamma_{bac}\right) \quad,   \\ 
												&=& -e^c_{~\nu}e^{b\mu}\gamma^a\Gamma_{abc} \quad,  
\end{IEEEeqnarray*} 
where we have used $\Gamma_{abc}=-\Gamma_{bac}$ in the derivation.  Substituting this result back into equation~\eqref{eq:DsmudtauGamma} we see that $\frac{\mathrm{D}S^\mu}{\dd\tau}=0$, which means that $S^\mu$ is parallel transported along the geodesic. This result verifies our claim in section~\ref{sec:motionspin}.

\section{Convention of elliptic integrals}
\label{sec:ellipticfunction}
The incomplete elliptic integrals of first and third kind are given by~\cite{Gradshteyn:1996table}
\begin{IEEEeqnarray}{rCl}
	\mathrm{F}(\phi,m) &=& \int_0^\phi \frac{\dd\theta}{\sqrt{1-m\sin^2\theta}} \quad,  \\ 
	\mathrm{\Pi}(n,\phi,m)&=& \int_0^\phi \frac{1}{1-n\sin^2\theta}\frac{\dd\theta}{\sqrt{1-m\sin^2\theta}} \quad,  
\end{IEEEeqnarray} 
and the corresponding complete ones are 
\begin{IEEEeqnarray}{rCl}
	\mathrm{K}(m) &=&\mathrm{F}\left(\frac{\pi}{2},m\right) =\int_0^\frac{\pi}{2} \frac{\dd\theta}{\sqrt{1-m\sin^2\theta}} \quad,  \\ 
	\mathrm{\Pi}(n,m)&=& \mathrm{\Pi}\left(n,\frac{\pi}{2},m\right)=\int_0^\frac{\pi}{2} \frac{1}{1-n\sin^2\theta}\frac{\dd\theta}{\sqrt{1-m\sin^2\theta}} \quad.  \label{eq:completePidef}
\end{IEEEeqnarray} 

\section{Asymptotic behaviour for the complete elliptic integral of the third kind $\Pi(n,m)$}

In this appendix we consider the asymptotic behaviour of the complete elliptic integral $\mathrm{\Pi}(n,m)$ when both $n$ and $m$ approaches to $1$. Since both of $n=1$ and $m=1$ are singularities of $\mathrm{\Pi}(n,m)$, the divergent behaviour of it depends on the path of $(n,m)$ approach to $(1,1)$. 
For our purpose, it's sufficient to consider the behaviour of $\Pi(1-z\delta,1-\delta)$ for $\delta\to0$. By definition~\eqref{eq:completePidef}, we have 
\begin{IEEEeqnarray}{rCl}
	\mathrm{\Pi}(1-z\delta,1-\delta) &=& \int_0^\frac{\pi}{2} \frac{1}{1-(1-z\delta)\sin^2\theta}\frac{\dd\theta}{\sqrt{1-(1-\delta)\sin^2\theta}} \quad,  \nonumber \\ 
	&=& \int_0^\frac{\pi}{2} \frac{1}{\cos^2\theta+z\delta\sin^2\theta}\frac{\dd\theta}{\sqrt{\cos^2\theta+\delta\sin^2\theta}} \quad,  \nonumber \\ 
	&=& \int_0^\frac{\pi}{2} \frac{1}{\sin^2\theta+z\delta\cos^2\theta}\frac{\dd\theta}{\sqrt{\sin^2\theta+\delta\cos^2\theta}} \quad.  \label{eq:Pideltaint}
\end{IEEEeqnarray} 
In the last line, we have changed the variable from $\theta$ to $\frac{\pi}{2}-\theta$. The integral in equation~\eqref{eq:Pideltaint} is singular in the limit $\delta\to0$, and the main contribution comes from the vicinity of $\theta=0$. Therefore, to the leading order, we can expand the integrand near $\theta=0$ and extending the integration range to be $(0,\infty)$. Then for $\delta>0$ and $\mathrm{Re}(z)>0$ we obtain 
\begin{IEEEeqnarray}{rCl}
	\mathrm{\Pi}(1-z\delta,1-\delta)&\approx&\int_0^\infty \frac{1}{\theta^2+z\delta}\frac{\dd\theta}{\sqrt{\theta^2+\delta}}  \quad, \nonumber \\ 
									&=& \frac{\cosh^{-1}(\sqrt{z})}{\sqrt{(z-1)z}}\frac{1}{\delta} \quad,  \label{eq:Picosh}
\end{IEEEeqnarray} 
where $\cosh(z)$ denotes the hyperbolic cosine of $z$.
Since $\mathrm{\Pi}(1-z\delta,1-\delta)$ and $\mathrm{\Pi}(1-z^*\delta,1-\delta)$ are complex conjugated, we have 
\begin{IEEEeqnarray}{rCl}
	\mathrm{\Pi}(1-z\delta,1-\delta)+\mathrm{\Pi}(1-z^*\delta,1-\delta) &=& 2\mathrm{Re}\left(\frac{\cosh^{-1}(\sqrt{z})}{\sqrt{(z-1)z}}\right)\frac{1}{\delta} \quad. \label{eq:PiPistarsum}
\end{IEEEeqnarray} 

\section{The critical behaviour of non-relativistic roots in \acrshort{rn} spacetime}
\label{sec:rootexpansion}

As we have mentioned, for the gravitational lensing of a particle in spherical symmetrical spacetime, there would be a critical value $\eta_c$ for $\eta$ ($\eta$ is defined in equation~\eqref{eq:uetadef} using impact parameter and speed), such that when $\eta\to\eta_c$, particles will wind around the black infinitely many times~\cite{Bozza:2002zj}. Corresponding to $\eta_c$, the first two positive roots $u_2$ and $u_3$ of $P_{RN}(u)$ in equation~\eqref{eq:upolyRN} will become equal, $u_2=u_3=u_c$. Both $\eta_c$ and $u_c$ can be determined by solving equation $P_{RN}=0$ and $P_{RN}'=0$ for $\eta$ and $u$. Setting $v=0$ we obtain 
\begin{IEEEeqnarray}{rCl}
	P_{RN}(u)=0\quad  \Rightarrow\quad &-u^4+\frac{2}{h^2}u^3-\left(\frac{1}{\eta^2}-\frac{1}{h^2}\right)u^2+\frac{2}{\eta^2h^2}u=0&\quad, \label{eq:prnv0}\\ 
	P_{RN}'(u)=0 \quad  \Rightarrow\quad &-4u^3 +\frac{6}{h^2}u^2-2 \left(\frac{1}{\eta^2}-\frac{1}{h^2}\right)u=0&\quad, \label{eq:prnpv0}
\end{IEEEeqnarray} 
from which we can eliminate $\eta$ to get
\begin{IEEEeqnarray}{rCl}
	h^4 u^4 -4 h^2 u^3 +4u^2 -u &=& 0 \quad, \label{eq:uceqnoeta} 
\end{IEEEeqnarray} 
Quartic equation~\eqref{eq:uceqnoeta} has four roots $0=u_1<u_2=u_3<u_4$. We see that $u_c=u_2=u_3$ is a double root\footnote{This is what one would expect because to get equation~\eqref{eq:uceqnoeta} we demand $\eta$ equals to the critical value $\eta_c$.}, while $u_4$ is the largest root of equation~\eqref{eq:uceqnoeta}.

For $v=0$, $\eta_c$ can be determined using equation~\eqref{eq:prnpv0}~\cite{Pang:2018jpm}
\begin{IEEEeqnarray}{rCl}
	\eta_c &=& \sqrt{\frac{ 1-h^2 u_c}{u_c \left(2 h^2 u_c^2-3 u_c+1\right)}} \quad. \label{eq:etacrn}
\end{IEEEeqnarray} 
When there is a small deviation from $\eta_c$ such that 
\begin{IEEEeqnarray*}{rCl}
	\eta &=& \eta_c+\delta^2 \quad, 
\end{IEEEeqnarray*} 
The roots $u_2$ and $u_3$ of the equation~\eqref{eq:prnv0} can be expressed as small deviations from $u_c$. In fact, we have~\cite{Hinch:1991pm} 
\begin{IEEEeqnarray}{rCl}
	u_2 &=& u_c-\sqrt{\alpha}\delta \quad, \quad u_3=u_c+\sqrt{\alpha}\delta \quad. \label{eq:u2u3alphadelta}
\end{IEEEeqnarray} 
Putting $u_2$ and $u_3$ given by~\eqref{eq:u2u3alphadelta} into the quartic equation~\eqref{eq:prnv0} of $P_{RN}(u)=0$, we see that terms of order $\delta^0$ and $\delta^1$ vanishes automatically, and demanding the $\delta^2$ term to vanish we can solve out $\alpha$

\begin{IEEEeqnarray}{rCl}
	\alpha &=& \frac{2u_c^2(1-h^2u_c)(1-2u_c+h^2u_c^2)}{\eta_c(4h^4u_c^3-9h^2u_c^2+6u_c-1)} \quad. \label{eq:alphaRNv0}
\end{IEEEeqnarray} 
For $v=0$, equation~\eqref{eq:prnv0} always has a root $u_1=0$ no matter whether $\eta$ deviates from $\eta_c$, while the largest root $u_4$ deviates from the root of equation~\eqref{eq:uceqnoeta} only at order of $\delta^2$\footnote{In contrast to $u_2$ and $u_3$, whose deviation are of order of $\delta$, as we can see from equation~\eqref{eq:u2u3alphadelta}.}, and won't contribute to the leading order expansion of $\chi_0$~\cite{Pang:2018jpm}. Therefore, in the expansion~\eqref{eq:chi0RNexp}, we can set $u_4$ to be the largest root of the quartic equation~\eqref{eq:uceqnoeta}.

For Schwarzschild case, letting $h=0$ in equation~\eqref{eq:prnpv0} we can see that $u_c=\frac{1}{4}$, which result in $\eta_c=4$ and $\alpha=\frac{1}{32}$ as we have used in section~\ref{sec:exampleSch}.


\nocite{*}
\input{./parallelspin_v5.bbl}

\end{document}

%% file: parallelspin_v5.bbl
\providecommand{\href}[2]{#2}\begingroup\raggedright\endgroup